\documentstyle[12pt]{article}
\def\pj{\hspace{-.26cm}}
\def\fpj{\hspace{-.7cm}}
\def\partsl{\partial\hspace{-2mm}/}
\def\asl{\bar{A}_a\hspace{-3.5mm}/}
\def\thalf{{\textstyle{\frac{1}{2}}}}
\newcommand{\vm}[1]{\mbox{\bf#1}}
\newcommand{\be}{\begin{eqnarray}}
\newcommand{\ee}{\end{eqnarray}}
\newcommand{\beq}{\begin{equation}}
\newcommand{\eeq}{\end{equation}}

\title{\begin{flushright}
{\normalsize NUC-MINN-98/13-T\\
December 1998 \\}
\end{flushright}
\vspace*{0.2in}
{\bf LOW-ENERGY THEOREMS FOR QCD AT FINITE TEMPERATURE AND CHEMICAL
POTENTIAL}}
\author{{\bf I. A. Shushpanov$^{1,2}$, J. I. Kapusta$^{1}$ and
P. J. Ellis$^{1}$} \vspace*{0.1in} \\
{\it $^{1}$School of Physics and Astronomy,
University of Minnesota}\\
{\it Minneapolis, MN 55455, USA} \vspace*{0.1in} \\
{\it $^{2}$Institute for Theoretical and Experimental Physics}\\
{\it B. Cheremushkinskaya 25, Moscow 117259, Russia} }
\date{~}

\begin{document}

\maketitle

\begin{abstract}
The low energy theorems for QCD are
generalized to finite temperature and chemical potential,
including non-zero quark masses.
\end{abstract}

\vspace*{1.0in}


\newpage

Low-energy theorems for quantum chromodynamics (QCD) at zero temperature
and density were derived long ago by Novikov {\it et al.} \cite{SVZ+N}.
They are useful in a number of contexts, for instance in constraining
effective theories or in assessing lattice gauge calculations.
Recently the theorems were generalized to finite temperature ($T>0$) for the
pure glue sector of QCD \cite{KET}. The purpose of the present note is
to provide a further generalization by including the quark sector
and allowing for non-zero density or chemical potential, $\mu$.

For clarity we shall consider one quark flavor since the case of several
flavors is trivially obtained by introducing flavor-dependent masses, $m_0$,
and chemical potentials, $\mu$, and appropriately summing over the flavor
index. In the imaginary time approach the partition function
takes the familiar form
\be
Z&\pj=&\pj\int [d \bar{A}] [dq] [d\bar{q}] \nonumber\\
&&\fpj\times\exp\left\{\int\limits_0^{1/T} d\tau d^3 x
\left[-\frac{1}{4 g_0^2} \bar{F}^{\mu\nu}_a\bar{F}_{\mu\nu}^a
+\bar{q}(i\partsl-\thalf\asl\;\lambda^a+\mu \gamma_0-m_{0})q
\right]\right\}\!.\label{Z}
\ee
Here we have suppressed the gauge fixing and Fadeev-Popov ghost terms, as
well as the quark color labels since they are inessential here. The
generators of color $SU(3)$ are denoted by $\lambda^a$, and
the gluon fields and field strength tensors have been
scaled by the bare coupling constant $g_0$: $\bar{A}_a^\mu=g_0{A}_a^\mu$
and $\bar{F}^a_{\mu\nu}=g_0{F}^a_{\mu\nu}$.

We first consider the case where the quark mass is set to zero.
The grand potential of the system is defined in the usual way,
$\Omega = -T \ln Z$, and we have
\beq
\frac{\partial }{\partial (-1/g_0^2)} \frac{\Omega}{V} =
-\frac{g_0^2}{4}
\left\langle F_a^{\mu\nu}(0,{\vm 0}) F^a_{\mu\nu}(0,{\vm 0})\right \rangle
\equiv-\frac{g_0^2}{4}\left\langle F^2\right \rangle\;,\label{gder}
\eeq
where $V$ is the volume of the system. The angle brackets denote a thermal
average.

This derivative can be calculated in another way by
using an explicit form for $\Omega/V$ determined on dimensional grounds.
Regulation of the ultraviolet divergences of the theory
introduces a mass scale, $M$, at which the subtractions are performed.
The results can be written in terms of the non-perturbative
dimensionful parameter
\beq
\Lambda =M \exp\left\{\int\limits^\infty_{\alpha_s (M)}
\frac{d\alpha_s}{\beta_s (\alpha_s)} \right\}\;,
\eeq
where $\alpha_s=g_0^2/4\pi$ and $\beta_s$ is the Gell-Mann-Low function:
$M d\alpha_s/dM = \beta_s$.
There are two additional dimensionful parameters, namely $T$ and $\mu$.
Since the grand potential is an observable quantity with
zero anomalous dimension \cite{larry}, we can write in general
\beq
\frac{\Omega}{V} = \Lambda^4 f \left(\frac{T}{\Lambda},
\frac{\mu}{\Lambda} \right )\;, \label{odim}
\eeq
where $f$ is some function. We note that at zero temperature and
chemical potential a form proportional to $\Lambda^4$ can be formally 
justified within a well-defined regularization scheme \cite{SVZ+N}.
In Eq. (\ref{odim}) $g_0$ is involved only
through $\Lambda$, hence we obtain
\beq
\frac{\partial }{\partial (-1/{g_0^2})} \frac{\Omega}{V} =
-\frac{4\pi \alpha_s^2}{\beta_s (\alpha_s)}
\left (4-T \frac{\partial }{\partial T}
-\mu\frac{\partial }{\partial \mu} \right ) \frac{\Omega}{V}\;.
\eeq
This leads to the chain of equations
\beq
\left ( 4-T \frac{\partial }{\partial T}
-\mu\frac{\partial }{\partial \mu} \right ) \frac{\Omega}{V}=
\frac{\beta_s (\alpha_s)}{4\alpha_s}\left\langle F^2\right \rangle
=\left\langle\theta_\mu^\mu(0,\vm{0})\right\rangle
= {\cal E} -3 P \;.\label{thrm}
\eeq
We have used the standard QCD result \cite{Tr} for the trace of the
improved energy-momentum tensor density
$\theta_\mu^\mu$ \cite{cal70}.
We have also used the standard thermodynamic relation for the energy density
\beq
{\cal E} = \left(1 -T \frac{\partial }{\partial T}
-\mu \frac{\partial }{\partial \mu} \right)\frac{\Omega}{V}\;,
\eeq
and the pressure $P=-\Omega/V$.

We turn now to the case where the quark mass is non-zero.
Since $\Omega$ is a physical observable it has to be expressed in terms
of renormalization group invariant quantities.
However, the quark mass has anomalous dimension
and depends on the scale $M$. The renormalization group equation
for $m_0(M)$, the running mass, is $(M/m_0)dm_0/dM = -\gamma_m$
and we use the $\overline{\rm MS}$ scheme for which $\beta_s$ and 
$\gamma_m$ are independent of the quark mass \cite{muta}.
Upon integration the renormalization group invariant mass is given by
\beq
\hat{m}=m_0(M) \exp\left\{\int\limits^{\alpha_s(M)}
\frac{\gamma_m (\alpha_s)}{\beta_s (\alpha_s)} d \alpha_s \right\}\;,
\eeq
where the indefinite integral is evaluated at $\alpha_s(M)$.
Then Eq. (\ref{odim}) becomes
\beq
\frac{\Omega}{V} = \Lambda ^4 h \left (\frac{T}{\Lambda},
\frac{\mu}{\Lambda}, \frac{\hat{m}}{\Lambda} \right )\;, \label{om}
\eeq
where $h$ is some function. Proceeding as before we obtain
\be
\left ( 4-T \frac{\partial }{\partial T}
-\mu \frac{\partial }{\partial \mu} \right ) \frac{\Omega}{V}&\pj=&\pj
\frac{\beta_s (\alpha_s)}{4\alpha_s}\left\langle F^2 \right\rangle +
\left [1+\gamma_m (\alpha_s) \right ] m_0
\left\langle{\bar q}{q}\right\rangle\nonumber\\
&\pj=&\pj\left\langle\theta_\mu^\mu(0,\vm{0})\right\rangle
= {\cal E} -3 P \;. \label{mum}
\ee
Here we have used the trace of the energy-momentum tensor for QCD with
quarks \cite{Tr,ji}. Clearly the physical quantities in Eq. (\ref{mum})
must obey the same relation as in Eq. (\ref{thrm}).

We can iterate this procedure by taking $n$ further derivatives of Eq.
(\ref{mum}). In doing so it is convenient to note that since
the grand potential density and its derivatives are independent of the
scale $M$ at which the ultraviolet divergence are regulated, we can
choose any scale to prove the result. It is convenient to pick a
sufficiently large scale that we can take the lowest order expressions,
$\beta_s(\alpha_s)\rightarrow-b\alpha_s^2/2\pi$, where $b=(11N-2N_f)/3$
for the $SU(N)$ gauge group with $N_f$ flavors, and
$(1+\gamma_m)\rightarrow1$. Then it is straightforward
to obtain the following relation
\begin{eqnarray}
&&\fpj(-1)^n \left (4-T \frac{\partial }{\partial T}
-\mu\frac{\partial }{\partial \mu} \right )^{n+1} \frac{\Omega}{V}=
\left ( T \frac{\partial }{\partial T}
+\mu \frac{\partial }{\partial \mu} -4\right )^n
\left\langle \theta_\mu^\mu \right\rangle \nonumber\\
&&=\int d\tau_n d^3 {x_n} \cdots \int d\tau_1 d^3 x_1
\left\langle \theta_\mu^\mu (\tau_n,\vm{x}_n)\cdots
\theta_\mu^\mu (\tau_1,\vm{x}_1)
\theta_\mu^\mu (0,\vm{0})\right\rangle_c\;. \label{ntrace}
\end{eqnarray}
Here the subscript $c$ means that only the connected diagrams are to be
included and the limits of the imaginary time integrations are suppressed.
It is possible to perform a similar analysis for a renormalization
group invariant operator ${\cal O}$
\begin{eqnarray}
&&\fpj\left ( T \frac{\partial }{\partial T}
+\mu\frac{\partial }{\partial \mu} - d \right )^n
\left\langle {\cal O}\right\rangle \nonumber\\
&&=\int d\tau_n d^3 x_n \cdots \int d\tau_1 d^3 x_1
\left\langle \theta_\mu^\mu (\tau_n,\vm{x}_n)\cdots
\theta_\mu^\mu (\tau_1,\vm{x}_1)
{\cal O}(0,\vm{0})\right\rangle_c\,, \label{opr}
\end{eqnarray}
where $d$ is the canonical mass dimension of ${\cal O}$.
If the operator ${\cal O}$ also has anomalous dimension then the
corresponding $\gamma$-function will have to be included.
Equations (\ref{ntrace}) and (\ref{opr}) differ from those found
previously \cite{KET} by the addition of the operator
$\mu\partial/\partial \mu$. One could also similarly generalize the
finite-momentum low-energy theorems \cite{KET,ms}.

As an example, where we can evaluate the left hand side of
Eq. (\ref{ntrace}), consider the case where $\mu$ and/or $T$ is much
greater than $\Lambda$ so that perturbation theory is valid.
Considering the first term in the $\beta_s$-function, the behavior
of the strong coupling constant as a function of chemical potential
and temperature can be written
\beq
\label{alpha}
\alpha_s(\xi)=\frac{2\pi}{b\ln (\xi/\Lambda)}\;,
\eeq
where $\xi=Ty(\mu/T)$ and we need not specify the function $y$.
The perturbative expression for the pressure in $SU(N)$ gauge theory up
to two-loop order is $\cite{Kap}$
\begin{eqnarray}
P&\pj=&\pj\frac{\pi^2}{45}(N^2-1) T^4 +
N\left ( \frac{7\pi^2 T^4}{180} +\frac{\mu^2 T^2}{6} +
\frac{\mu^4}{12\pi^2}\right ) \nonumber\\
&&\qquad-\frac{\pi (N^2-1) }{144}\alpha_s (\xi)
\left ( (4N+5)T^4 +\frac{18}{\pi^2}\mu^2 T^2 +\frac{9}{\pi^4}\mu^4
\right )\,.\label{omega}
\end{eqnarray}
Then from Eqs. ($\ref{alpha}$) and ($\ref{omega}$) we find
\begin{eqnarray}
&&\fpj\left ( T \frac{\partial }{\partial T}
+\mu\frac{\partial }{\partial \mu} -4\right )^n
\left \langle \theta_\mu^\mu\right\rangle =\frac{b}{288}(N^2-1) \nonumber\\
&& \times\alpha^{n+2}_s (\xi)
\left ( \frac{-b}{2\pi}\right )^n \!(n+1)!
\left ( (4N+5)T^4 + \frac{18}{\pi^2} \mu^2 T^2 + \frac{9}{\pi^4} \mu^4
\right )\!. \label{Pert}
\end{eqnarray}
This provides a requirement on model or lattice calculations of the
right hand side of Eq. (\ref{ntrace}).
Note, however, that if we consider the next order in the
$\beta_s$-function then, for $N=3$ and one flavor in the
$\overline{{\rm MS}}$-scheme, we obtain a correction
factor of order $(1+1.4 \alpha_sn)$ to Eq. ($\ref{Pert}$).
Thus, at this level of approximation for
$P$ and $\alpha_s$, an arbitarily large number of derivatives cannot
be taken without at some point losing all accuracy.

In conclusion we have shown that at finite temperature and chemical
potential the low-energy theorems of Novikov
{\it et al.} \cite{SVZ+N}  hold
provided that the operators $T\partial/\partial T$ and
$\mu\partial/\partial \mu$ are appropriately included
as in Eqs. (\ref{ntrace}) and (\ref{opr}).

I.A.S. acknowledges the warm hospitality extended to him during his stay
at the University of Minnesota.
This work  was supported in part by the following grants: CRDF RP2-132,
Schweizerischer National Fonds 7SUPJ048716 and US Department
of Energy DE-FG02-87ER40328.

\end{document}